\documentclass{icrc}

\usepackage{amssymb}
\usepackage{times}
\usepackage{graphicx} % when using Latex and dvips
\begin{document}

\title{Tau Neutrinos from Astrophysical and Cosmological Sources}
\author[1]{Jane H. MacGibbon}
\affil[1]{Code SN3, NASA Johnson Space Center Houston Texas 77058 USA}
\author[2]{Ubi F. Wichoski}
\affil[2]{Depto. de F\'{\i}sica, CENTRA-IST, Av. Rovisco Pais, 1 - Lisbon \
1049-001 Portugal}
\author[3]{Bryan R. Webber}
\affil[3]{Cavendish Laboratory, University of Cambridge, England}
\correspondence{jane.macgibbon1@jsc.nasa.gov}

\firstpage{1}
\pubyear{2001}

% \titleheight{11cm} % uncomment and adjust in case your title block
                     % does not fit into the default and minimum 7.5 cm

\maketitle

\begin{abstract}
Previous work on the neutrino spectra from high energy sources has not 
included the tau neutrinos directly produced by the decays in the source. 
Here we consider the tau neutrino component and discuss how its inclusion 
modifies the expected neutrino spectra. We discuss implications for 
interpreting any observed tau neutrino component in TeV - UHE events as 
evidence of $\nu_{\mu} \rightarrow \nu_{\tau}$ oscillations.
\end{abstract}

\section{Introduction}

The observation of high energy neutrinos from beyond Earth will open a new 
window on astrophysics and cosmology. Presently much effort is being 
expended on designing and deploying detectors and methodologies capable of 
resolving the flavour of the incoming neutrino flux. One motivation for 
doing this is that the absolute sensitivity, or relative sensitivity 
compared with background, may be significantly greater for one neutrino 
species than the other species \citep{DRS}. Another primary motivation 
for identifying the neutrino flavour is to test the hypothesis that 
neutrinos undergo flavour oscillation as they propagate from source to 
Earth. Indeed a number of atmospheric, accelerator and solar neutrino 
experiments hint at the existence of neutrino oscillations, although the 
experiments are not yet sufficiently consistent to delineate a unique 
solution \citep{Akhm}. The most favoured solution, fitting 
the latest SuperKamiokande data, involves 
$\nu_{\mu} \rightarrow \nu_{\tau}$ oscillation with a mass 
difference of $\Delta m_{\nu}^2 \sim 10^{-3} \mbox{eV}^2$. Because 
the non-oscillation production of $\nu_{\tau}$ is assumed to be 
negligible relative to $\nu_{e}$ and $\nu_{\mu}$ production, and 
the cascade $\nu_{\tau}$ produced by collisions of primary $\nu_{e}$ and 
$\nu_{\mu}$ flux with the relic cosmic neutrino background is orders of 
magnitude less than the primary $\nu_{e}$ and $\nu_{\mu}$ flux, any 
detected $\nu_{\tau}$ component from a source outside 
the Solar System is currently expected to be indicative of oscillation. 

In this paper, we briefly report some results of our investigation into 
high energy $\nu_{\tau}$ production, presented elsewhere \citep{MWW}. 
We illustrate that the $\nu_{\tau}$ component at the high energy end 
in the hadronic decays is significantly higher than previously assumed 
and that this can have observational consequences.

\section{Neutrino Production}

A number of sources for TeV-UHE neutrinos have been postulated: for 
example active galactic nuclei (AGN) \citep{Ath,Hus}; topological 
defects such as superconducting, ordinary or VHS cosmic 
strings \citep{BS,HSW,WMB}; supermassive gauge and scalar particle (X particle) 
decay or annihilation \citep{AHK}; and Hawking evaporation of primordial 
black holes \citep{MC,MW,Hal}. In all 
of these scenarios, the final state particle distributions 
are expected to be dominated by hadronic decays at the source. In these 
decays over 90\% of the final state emission will be $\pi^{0}$, 
$\pi^{+}$, and $\pi^{-}$, with the remainder mainly nucleons which 
decay into protons and antiprotons. On astrophysical timescales, 
the $\pi^{\pm}$ decay into $\stackrel{\mbox{\tiny (-)}}{\nu_{e}}$ and 
$\stackrel{\mbox{\tiny (-)}}{\nu_{\mu}}$ and $e^{\pm}$. It is 
known that the final cluster states (pions and nucleons) in QCD jets at 
accelerator energies are well described by the fragmentation function
\begin{equation}
\label{primdecay}
{{d N} \over {d x}} \, = \, {{15} \over {16}} x^{-3/2} (1 - x)^2 \, , 
\end{equation} 
where $x= E/m_J$ and $m_J$ is the total energy of the decaying jet 
\citep{HSW}. This distribution continues down to $E \sim 1$ GeV. When 
convolved with the $\pi^{\pm}$ decay, Eq.(\ref{primdecay}) leads to, 
similarly, a dominant $E^{-3/2}$ term in the $\nu_{e}$  and 
$\nu_{\mu}$ spectra at $x \lesssim 0.1$ and 
$d N_{\nu_{e},\nu_{\mu}}/dE \rightarrow 0$ as $x \rightarrow 1$. 
(For a full derivation of the $\nu_{e}$ and 
$\nu_{\mu}$ spectra using fragmentation function (\ref{primdecay}) 
see \citep{WMB}.) 

The production channels for the $\nu_{\tau}$, however, are substantially 
different. The tau neutrinos are only produced in significant numbers 
once the tau lepton and heavy quark masses, $m_{\tau} = 1.78$ GeV, 
$m_b \sim 4$ GeV, $m_t \sim 174$ GeV, are surpassed. The greatest 
contribution comes from the $t$ quark decay. 
Because $\nu_{\tau}$ production is suppressed compared with other species, 
it has been assumed in flux calculations that the $\nu_{\tau}$ spectrum is 
orders of magnitude less than the $\nu_{e}$ and $\nu_{\mu}$ spectra at 
all $x$. This is not so. While indeed the total number of $\nu_{\tau}$ 
produced per jet is less than $10^{-3}$ of the total number of 
$\nu_{e}$ and $\nu_{\mu}$, the high $x$ tau neutrinos are 
predominantly produced by the initial decays of the heavier quarks with 
shorter lifetimes and the $\nu_{e}$ and $\nu_{\mu}$ are produced by the 
final state cluster decays of the much lighter pions. This leads to 
significantly greater relative contribution from $\nu_{\tau}$ at high 
$x$ than previously assumed. The fragmentation distribution 
(\ref{primdecay}) is no longer relevant for the tau neutrino.

In Fig.(1) we show the $\nu_{\tau}$ spectrum, together with the 
$\nu_{e}$ and $\nu_{\mu}$ spectra, generated by the decay of 10 TeV 
$q \bar{q}$ jets. To simulate these spectra we used the QCD event 
generator HERWIG \citep{Cetal} and the process 
$e^+ e^- \rightarrow q_i \bar{q}_i \; (g)$, $i$ = all $q$ flavours. 
Consistent spectra are obtained with 
PYTHIA/JETSET (see \cite{MWW}). Note that in the region 
$0.1 \lesssim x \lesssim 1$, the tau neutrinos make up more than one 
tenth of the total neutrino contribution. We also note that below 
$x \lesssim 0.1$, $d N_{\nu_{\tau}}/d E$ falls off with roughly an 
$E^{-1/2}$ slope, and not the $E^{-3/2}$ slope of the 
$\nu_{e}$ and $\nu_{\mu}$ spectra.
\begin{figure}[t]
\label{fig1a}
\includegraphics[width=6.0cm,angle=-90]{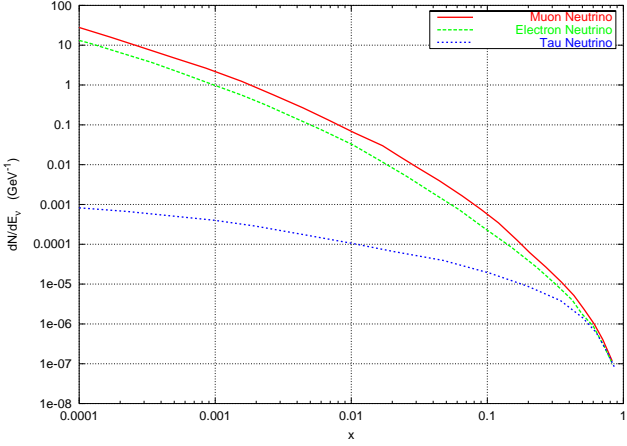}
\caption{The neutrino spectra, $d N/d E$ vs $x=E/m_J$, generated by 
the decay of $m_J=10$ TeV $q\bar{q}$ jets. The solid line represents 
$\nu_{\mu}$, the dashed line represents $\nu_{e}$, and the dotted line 
represents $\nu_{\tau}$.}
 \end{figure}
Because the 1 GeV - UHE cosmic ray flux backgrounds are expected to fall 
off as $d N/d E \propto E^{-y}$ where $2 < y < 3$, we also plot 
$E^{2} d N/d E$ in Fig.(\ref{e2spect}).
\begin{figure}[t]
\label{e2spect}
 \includegraphics[width=6.0cm,angle=-90]{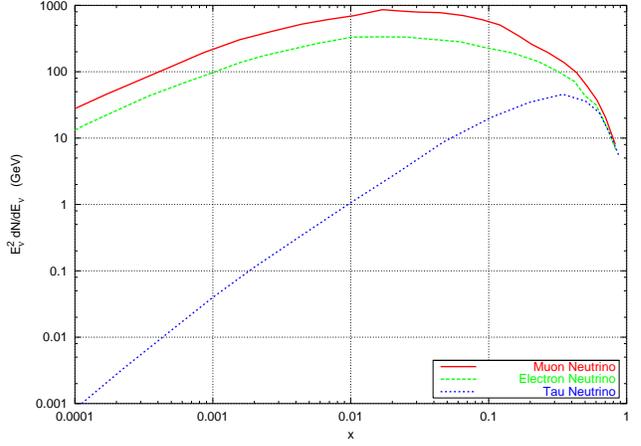}
 \caption{The $E^2 d N/d E$ vs $x=E/m_J$ neutrino spectra for the same 
jet decay as in Fig.(1).}
 \end{figure}

From our 300 GeV - 75 TeV simulations, we find that the $\nu_{\tau}$  
spectrum generated by $q \bar{q}$ jet decay can be parametrized as
\begin{eqnarray}
\label{fragtau}
{{d N_{\nu_{\tau}}} \over {d E_{\nu_{\tau}}}} &\simeq& 
\bigg(\frac{1}{2 m_J}\bigg) \Bigg[
\;0.15 \bigg(\frac{E_{\nu_{\tau}}}{m_J}\bigg)^{-1/2} - 
\; 0.36 \;+ \nonumber \\
& & 0.27 \bigg(\frac{E_{\nu_{\tau}}}{m_J}\bigg)^{1/2} - \;
0.06 \bigg(\frac{E_{\nu_{\tau}}}{m_J}\bigg)^{3/2} \Bigg] \; ,
\end{eqnarray}
per jet. Similarly the number (multiplicity) of neutrinos produced 
per $q \bar{q}$ jet and the average neutrino energy scale as
\begin{equation}
\label{multavetau}
N_{\nu_{\tau}} \simeq 0.035 \bigg(\frac{m_J}{GeV}\bigg)^{0.03}, \;
\bar{E}_{\nu_{\tau}} \simeq 0.2 \bigg(\frac{m_J}{GeV}\bigg)^{0.9} GeV,
\end{equation}
\begin{equation}
\label{multavee}
N_{\nu_{e}} \simeq 2.0 \bigg(\frac{m_J}{GeV}\bigg)^{0.3},  \;
\bar{E}_{\nu_{e}} \simeq 0.04 \bigg(\frac{m_J}{GeV}\bigg)^{0.7} GeV,
\end{equation}
\begin{equation}
\label{multavemu}
N_{\nu_{\mu}} \simeq 3.6 \bigg(\frac{m_J}{GeV}\bigg)^{0.3}, \;
\bar{E}_{\nu_{\mu}} \simeq 0.04 \bigg(\frac{m_J}{GeV}\bigg)^{0.7} GeV. 
\end{equation}
Note that the average $\nu_{\tau}$ energy is significantly higher that 
the average  $\nu_{e}$ and $\nu_{\mu}$ energies, consistent with our 
remarks above. We derive similar $d N_{\nu_{\tau}}/d E_{\nu_{\tau}}$, 
$N_{\nu_{\tau}}$ and $\bar{E}_{\nu_{\tau}}$ parametrizations for the 
$\nu_{\tau}$ generated by initial $t \bar{t}$ decay, $b \bar{b}$ decay, 
$\tau^{+} \tau^{-}$ decay, $W^{+} W^{-}$ decay and decays which include 
extensions to the Standard Model (e.g. SUSY and Higgs sectors) in 
\citep{MWW}.

In analogy with the use of the fragmentation function (\ref{primdecay}), 
we now apply (\ref{fragtau}), and its high energy extrapolation, to 
investigate the neutrino fluxes produced in astrophysical and cosmological 
scenarios.

\section{Neutrino Fluxes from Astrophysical and Cosmological Sources}

To calculate the expected $\nu_{\tau}$ flux from a given astrophysical or 
cosmological source, the above $\nu_{\tau}$ fragmentation function, 
Eq.(\ref{fragtau}), must be convolved with the initial distribution of 
the decaying particles, the Galactic or cosmological distribution of the 
sources and the redshift evolution and interactions of the emitted 
neutrinos as they propagate from source to Earth.
 
Here we briefly mention two scenarios. The new spectra for the tau neutrino 
fluxes expected from a number of other astrophysical and cosmological 
sources, including AGN and primordial black holes, are presented in our 
paper \citep{MWW}.

Our first example is the VHS cosmic string scenario \citep{WMB}. 
However, our remarks apply generically to $p=1$ cosmic string models. 
In the VHS scenario, 1 GeV - UHE particle fluxes are generated by the decay 
of supermassive scalar and gauge particles emitted by the long strings 
over the age of the Universe. 
The neutrino spectra expected at Earth from VHS strings 
with a mass per unit length of $G \mu = 10^{-8}$ are shown in 
Fig.(3) \citep{WMB}. 
The fraction of the $\nu_{\tau}$ component which is solely produced in the 
collisions of the primary $\nu_{e}$ and $\nu_{\mu}$ with the relic cosmic 
$1.9^oK$ neutrino background is represented by the dash-dotted line. 
(This is the only $\nu_{\tau}$ component presented in previous cosmic 
string papers.) The dotted line represents the $\nu_{\tau}$ directly 
produced in the hadronic decays of the string emission. As can be seen, 
our results give a significant increase in the expected $\nu_{\tau}$ 
signal at the highest energies. 
Note too that because the cosmic ray background falls off as $E^{-y}$, 
the highest energy region of the spectra is most relevant to detection. 
In Fig.(4), we show the neutrino spectra from a viable VHS model with 
$G \mu=10^{-10}$ and the comparisons with present and proposed detector 
sensitivities.

 \begin{figure}[t]
 \label{fluxescasc}
% \vspace*{2.0mm} % just in case for shifting the figure slightly down
 \includegraphics[width=6.0cm,angle=-90]{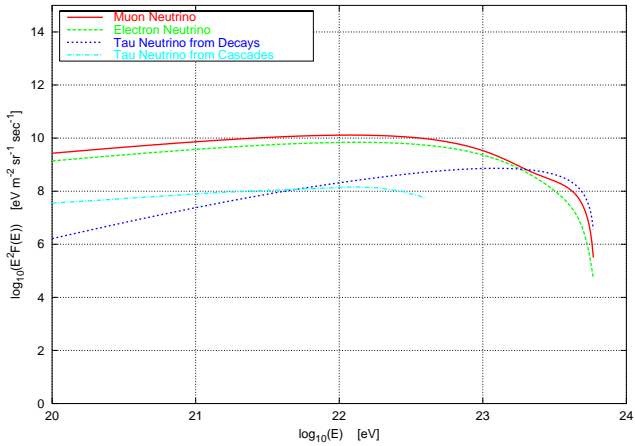} % .eps for Latex,
 \caption{The neutrino spectra from VHS cosmic strings with a mass 
per unit length of $G \mu=10^{-8}$. The dash-dotted  curve is the 
$\nu_{\tau}$ component produced by cascades off the relic neutrino 
background \citep{Yosh}.} 
 \end{figure}

As the second example, we briefly comment on the implications of our 
results for the simpzilla scenario. The annihilation of strongly 
interacting superheavy dark matter $X$ particles captured by our Sun 
(simpzillas) has recently been explored as a source of observable 
$\nu_{\tau}$ \citep{AHK}. In this scenario the $X$ particle decays into 
2 quarks or 2 gluons which then decay into jets. Because of the high 
solar density, it is argued that only the 
$t \rightarrow W \rightarrow \nu$ chain in the $X$ decay would 
produce a significant $\nu_{\tau}$ flux capable of escaping the Sun. 
Assuming the fragmentation function (\ref{primdecay}) applies to the $t$ 
quarks in this decay chain, the authors derive a distribution for the 
$\nu_{\tau}$ produced per $X$ particle decay which has a leading 
$E^{-3/2}$ slope above $m_t$. However, our full $W^{\pm}$ decay 
simulations show that the $\nu_{\tau}$ spectrum falls off with a much 
weaker slope, inconsistent with an $E^{-3/2}$ slope from the partial 
decay. As a resolution to this discrepancy, we believe that the 
application of Eq.(\ref{primdecay}) 
to the initial $t$ distribution from $X$ annihilation in the Sun is 
inappropriate because the short-lived t quarks are generated in the first 
step directly by the $X$ annihilation, whereas Eq.(\ref{primdecay}) is a 
fit to the final state cluster distributions. Thus the appropriate 
distribution of the $t$ quarks in this chain should be the initial 
distribution of the $X$ particles times the relevant branching ratio. 
This skews the expected $\nu_{\tau}$ distribution from solar $X$ 
annihilation to higher energies. The modifications are discussed in 
detail in \citep{MWW}.

 \begin{figure}[t]
 \label{fluxesdat}
% \vspace*{2.0mm} % just in case for shifting the figure slightly down
 \includegraphics[width=6.0cm,angle=-90]{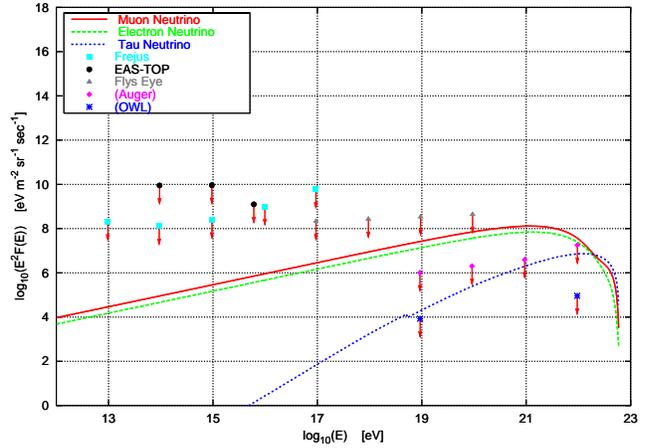} % .eps for Latex,
 \caption{The neutrino spectra from VHS cosmic strings with 
$G\mu=10^{-10}$. Points with arrows represent upper limits for the flux 
(taken from \citep{BS}).}
\end{figure}

\section{Conclusions}

Previous work has assumed that the $\nu_{\tau}$ component in TeV - UHE 
hadronic decays is negligible compared with the $\nu_{e}$ and $\nu_{\mu}$ 
components, at all neutrino energies. On examining QCD behaviour at 
accelerator energies and assuming similar behaviour continues to higher 
energies, we find this is not so. In particular, for neutrino energies 
in the decade below the energy of the decaying particle, the tau neutrino 
component is of similar magnitude to the $\nu_{e}$ and $\nu_{\mu}$ 
components. Below these energies the $\nu_{\tau}$ spectrum exhibits a 
slope of slightly less than $E^{-1/2}$  compared 
with the $E^{-3/2}$ slope of the $\nu_{e}$ and $\nu_{\mu}$ spectra. 
This analysis modifies the expected spectra in many astrophysical and 
cosmological high energy neutrino production scenarios. As a consequence, 
the observation of a significant $\nu_{\tau}$ to $\nu_{\mu}$ ratio 
at a given energy in high energy neutrino telescopes and detectors may 
be due to hadronic decay at the source and not 
$\nu_{\mu} \rightarrow \nu_{\tau}$ oscillation in transit. 

\begin{acknowledgements}
It is a pleasure to thank Mike Seymour for advice and Robert 
Brandenberger and Andrew Heckler for encouragement. 
UFW has been supported by ``Funda\c{c}\~ao para a 
Ci\^encia e a Tecnologia'' (FCT) under the program ``PRAXIS  XXI''.
\end{acknowledgements}

\end{document}